# Enabling an Anatomic View to Investigate Honeypot Systems: A Survey

Wenjun Fan, Zhihui Du, *Senior Member, IEEE,* David Fernández, and Víctor A. Villagrá

***Abstract*—A honeypot is a type of security facility deliberately created to be probed, attacked and compromised. It is often used for protecting production systems by detecting and deflecting unauthorized accesses. It is also useful for investigating the behaviour of attackers, and in particular, unknown attacks. For the past 17 years much effort has been invested in the research and development of honeypot based techniques and tools and they have evolved to become an increasingly powerful means of defending against the creations of the blackhat community. In this paper, by studying multiple honeypot systems, the two essential elements of honeypots - the decoy and the security program - are captured and presented, together with two abstract organizational forms - independent and cooperative - in which these two elements can be integrated. A novel decoy and security program (D-P) based taxonomy is proposed, for the purpose of investigating and classifying the various techniques involved in honeypot systems. An extensive set of honeypot projects and research, which cover the techniques applied in both independent and cooperative honeypots, is surveyed under the taxonomy framework. Finally, the taxonomy is applied to a wide set of tools and systems in order to demonstrate its validity and predict the tendency of honeypot development.**

***Index Terms*—Honeypots, Computer Security, Virtualization, Network Security, Intrusion Detection**

## I. Introduction

THE new domain of cyberspace is so pervasive that the US Department of Defense has put cyberspace on a par with land, sea, and air as a war-fighting domain [1]. Systems in cyberspace are constantly faced with cyber threats every day. In 2015, Symantec discovered 54 zero-day vulnerabilities, a 125 percent increase from the year before [2]. Since cyber threats cannot be eliminated completely, the strategy to securing cyberspace is to remove as many vulnerabilities as possible before they can be exploited [3]. A honeypot is a vital security facility aimed at sacrificing its resource to investigate unauthorized accesses in order to discover potential vulnerabilities in operational systems, and reduce the risks. Due to its unique design and application features, it can help to address the deficiencies of other existing security methods.

Firewalls are often deployed around the perimeter of an organization in order to block unauthorized access by filtering certain ports [4] and content, but they do little to evaluate the traffic. They can block all accesses to a certain service in order to prevent malevolent traffic, but this also blocks any benevolent traffic that wants to access the service. Conversely, honeypots are aimed at opening ports in order to capture as many attacks as possible for subsequent data analysis. An intrusion detection system (IDS) is used to evaluate the traffic and detect any inappropriate, incorrect, and anomalous activity. However, IDSs often have the “false alert problem”, i.e. signature (rule-based) IDSs often generate false negative alerts, whilst anomaly-based IDSs generate false positive alerts. Compared to an IDS, a honeypot has the big advantage that it never generates false alerts, because any observed traffic to it is suspicious since there is no production service running on the honeypot. Hence, an integration of a honeypot with an IDS can largely reduce the number of false alerts [5].

An intrusion prevention system (IPS), comprising a firewall plus an IDS, can evaluate the traffic and block malicious data. It acts as a shield against attacks, but it is not able to distinguish whether an application-layer request is normal or not. This drawback could potentially result in attacks permeating the shield without being detected. For example, a social engineering attacker may gain sensitive information by using a compromised legitimate username and password [6]. However, if an IPS integrates with a honeypot, the whole system can then capture all attacking activities regardless of whether they are performed by inside or outside adversaries. In addition, the data captured by honeypots can be used to create countermeasures, e.g. the automated intrusion response systems (AIRS) often uses honeypots as the data capture infrastructure [7].

Honeypots are often used to investigate currently-unknown attacks [5], [8]. The Blackhat community is intelligent enough to create new-unknown threats. A good way to investigate new threats is to capture the malicious activity step-by-step as it compromises a system. Honeypots therefore can add value to research by providing a sacrificial system to be attacked. Furthermore, it is worth observing what the adversaries do in the compromised system, such as communicating with other attackers and uploading new rootkits. Also, honeypots can effectively capture automated attacks [9], [10]. Due to the fact that automated attacks often target the entire network, honeypots can quickly capture them for investigation.

Hence, according to different security requirements, a variety of honeypots have been proposed, i.e. there is not only dedicated honeypot software [11], but also complex cooperative honeypot systems, such as honeynets [12] and hybrid systems [10], [9], [13], etc. However, there is a lack of a distinct method that can quickly catch the key points

W. Fan, D. Fernández and V.A.Villagrá are with the Department of Telematics Engineering, Technical University of Madrid, 28040, Madrid, Spain.
E-mail: efan@dit.upm.es, david@dit.upm.es, villagra@dit.upm.es.
Z. Du is with Tsinghua National Laboratory for Information Science and Technology, Department of Computer Science and Technology, Tsinghua University, 100084, Beijing, P.R.China.
E-mail: duzh@tsinghua.edu.cn

of the various honeypots and that can discover new insights, and advance research and development in this area. Our work proposes to address these problems. The main contributions of this paper can be summarized as follows:

- Two essential elements (decoy and security program) of a honeypot are captured. How these are organized is described, and this can provide a general view for analyzing diverse honeypot systems.
- A novel decoy and security program (D-P) based taxonomy is proposed to investigate different aspects of honeypot technology.
- Several development trends are identified by comparing honeypots according to the taxonomy.

The remainder of this article is organized as follows: Section 2 defines the core concepts and terminology; Section 3 proposes a way to investigate different honeypot technologies by providing a novel D-P based taxonomy; Section 4 surveys a number of honeypots based on the taxonomy in order to analyze their development; Section 5 makes a conclusion.

## II. Honeypot Anatomy

The first idea about honeypot comes from a book titled "The Cuckoo's Egg" [14] that described a series of events about tracking a hacker. The second material about honeypot was reported in a whitepaper [15]. The definition of honeypot was proposed by Spitzner [16]: "A honeypot is an information system resource whose value lies in the unauthorized or illicit use of that resource." However, it is more like a description of honeypot from its application value. We therefore provide a clear definition of what is a honeypot (see Figure 1).

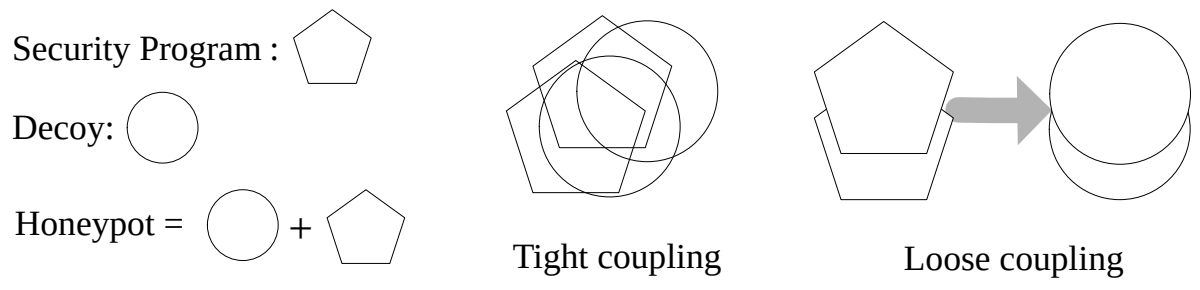

Fig. 1. Honeypot Anatomy: core elements and their organizations.

A **honeypot** includes two essential elements, decoys and security programs, and it is used to deliberately sacrifice its information resource allowing unauthorized and illicit usage for security investigation. The **decoy** can be any kind of information system resource, and the **security program** facilitates the security related functions, such as attack monitoring, prevention, detection, response and profiling. In addition, the security program should be running in stealthy to avoid being detected. Among the existing honeypot projects and honeypot research work, the terminology meanings were not consistent. Some refer to decoys as honeypots. For example, a decoy can be a fake digital entity. The terminology for digital entity acting as decoy is **honeytoken** [16]. In the book "The Cuckoo's Egg", Stoll deployed the honeytokens, i.e.,digital files, with security program to track a German hacker. Thus, the honeytoken is decoy, but Stoll's system is a honeypot system. Our definition clarifies that a vulnerable system without any security program is only a decoy rather than a honeypot, unless it is equipped with security program then we call it honeypot.

The organization of the two essential elements can be roughly categorised into two coupling degrees: loose and tight (see Figure 1). Coupling refers to the degree of direct knowledge that one component has of another. Loose coupling is one in which each component has, or makes use of, little or no knowledge of the definitions of other separate ones. It enables components to remain completely autonomous and unaware of each other while still interfacing with each. In contrast, tight coupling is when a group of components are highly dependent on one another, or they are built into the same impartible unit to perform the task. The **independent honeypot** refers to the one using tight coupling, and the **cooperative honeypot** indicates the one using loose coupling. Nawrocki et al. [11] surveyed a number of honeypot software that are independent honeypots, while complex systems such as the honeynets [12] and hybrid systems [10], [9], [13] are cooperative honeypots. In this paper, we use the term "honeypot" and "honeypot system" interchangeably.

## III. Review with D-P based Taxonomy

This section proposes a novel D-P based taxonomy as Figure 2 shows. The terminology is described by a technical approach, which can make their definitions distinct and easy to understand. The classification scheme are divided into two categories. The first category includes the features of a decoy, and the second one consists of the functions of a security program. The D-P based taxonomy is used as a basic conceptual model in order to investigate honeypot technology. Under this taxonomy framework, we review the typical honeypots and those specific honeypot-related techniques.

### A. Features of decoy

Decoy aims to capture data by being attacked. There are several primitive characteristics that can compose the design specification of a decoy.

*1) **Fidelity**:* It denotes the degree of exactness of an information system resource that the decoy provides to the attack. It is a criterion used to classify the interaction into three levels: low, medium, high (see Figure 3).

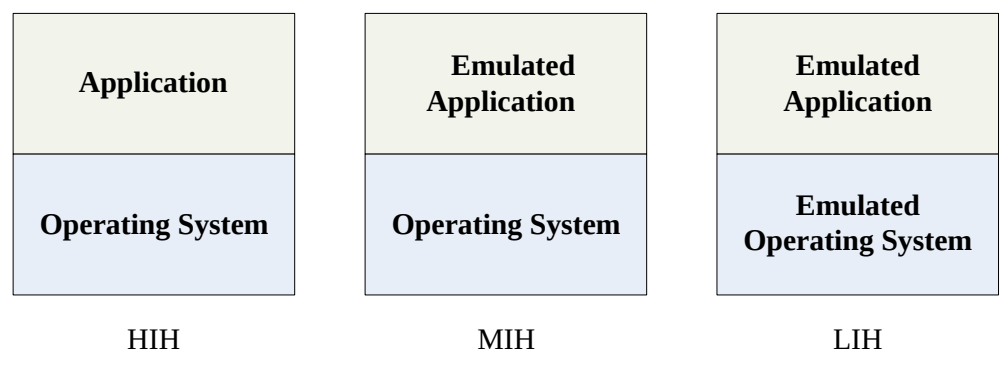

Fig. 3. Three types of fidelity.

**Low-interaction honeypot** (LIH) only provides a little interaction to adversaries. The decoy of LIH always has another name, facade. A traditional LIH, e.g. Honeyd [17], is a program emulating the protocols of operating systems with a limited subset of the full functionality. An adversary is not able to compromise a LIH because there is only OS fingerprinting artifice instead of real operating system on the LIH. A LIH system can provide security program to monitor the facade in order to capture the network activity.

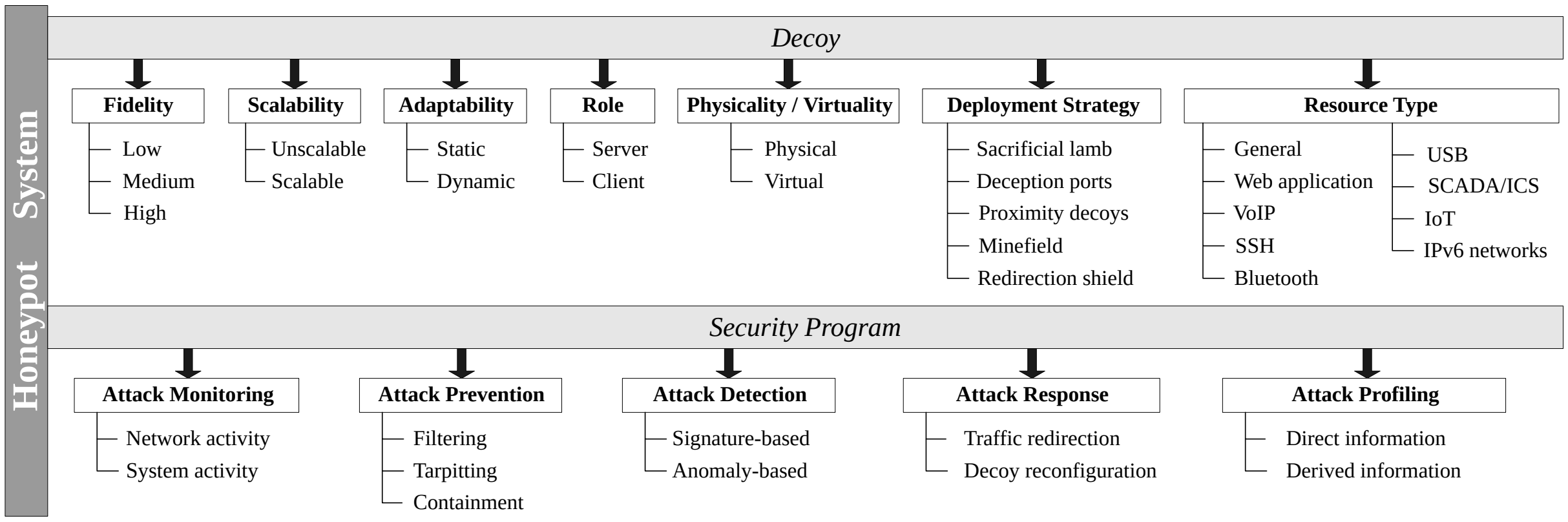

Fig. 2. D-P based Taxonomy of honeypot systems.

**Medium-interaction honeypot** (MIH) can provide much more interaction to the adversaries. However, unlike to LIH, MIH does not implement TCP/IP stacks by themselves. Instead, MIHs, e.g. Dionaea [18] and Cowrie [19], bind on sockets and leave the operating system do the connection management. In contrast with LIHs that implement network protocols, the simulation algorithm of MIHs is based on emulating logical application responses for incoming requests. Thus, the request arriving to the MIH will be watched and examined, and the fake responses will also be created by the security program of the MIH.

**High-interaction honeypot** (HIH) is fully functional system which can be completely compromised by adversaries. Its decoy is often a genuine system, such as Argos [8] and Cuckoo Sandbox [20]. Because the fully functional honeypot can be compromised, the HIH must equip security toolkits for system activity capture and outgoing traffic containment.

A **hybrid honeypot system** often consists of decoys of different interaction levels, e.g. Artail's hybrid honeypot framework [9], and Bailey's [10] and Lengyel's [13] hybrid honeypot architectures. In a hybrid system, the LIHs or MIHs are often used as front ends for large-scale deployment and the HIHs are used as back ends for deep investigation. Those distributed front ends are named **sinkholes**, which could be the devices (i.e. sensors, redirectors, etc.), such as network telescopes [21], darknet [22], blackholes [23], IMS [24], and iSinks [25], or software artifice assigned with a portion of routed IP address space. Instead of deploying a large number of HIHs across multiple networks, they can be centrally deployed in a consolidated location, which is called **honeyfarm**, such as the one used in Potemkin [26].

*2) **Scalability**:* it represents the capability to provide a growing amount of decoys, or its potential to be enlarged to accommodate that growth. It can be classified into two categories: unscalable and scalable. An unscalable honeypot only includes a certain number (one or more) of decoys and cannot change the number, e.g. Argos [8] can only monitor one virtual decoy. On the contrary, A scalable honeypot system can deploy multiple decoys and its security program is able to monitor those decoys simultaneously, e.g. Honeyd [17] is able to emulate multiple OS fingerprinting artifice at the same time. A honeynet is a type of scalable honeypot system. The term of **honeynet** was proposed by the work [12], [27], which defines that a honeynet is a network consisting of high interaction honeypots that provide real systems, applications, and services for adversaries to interact with. Besides, the data captured by the scalable honeypots deployed in multiple domains often need to be collected by secure channel and stored in an isolated data center for further analysis.

*3) **Adaptability**:* it refers to the reconfiguration capability to adapt the state of the decoy to changed circumstances. It has two levels: static and dynamic. Traditional static honeypots, e.g. Specter [28] and Dionaea [18], need the security researcher to determine the configuration beforehand and manually reconfigure it later. This static configuration scheme has several drawbacks: 1) it is a complex task to manually configure honeypots; 2) the static configuration scheme is not able to make an instant response to the intrusion event in time; 3) it is not able to adapt to the change of the objective cloned network timely. In contrast, a **dynamic honeypot** is able to timely adapt to the specific events. It is able to change the configuration periodically or even in real-time to adapt to the environment changes and respond to the intrusion events, e.g. Honeyd [17] and Glastopf [29].

*4) **Role**:* it describes in which side the decoy plays within a multi-tier architecture. A honeypot can play two roles: server and client. It refers to whether a honeypot actively detects malicious program or passively captures unauthorized traffic. Most honeypots are server side ones, e.g. Honeyd [17] and Dionaea [18], which passively wait being attacked. Adversaries find these honeypots on their own initiative and probe and attack them. Most server-side honeypots never advertise themselves, but some can "advertise" themselves, e.g. Glastopf [29] works like a normal web server with a number of ever requested vulnerable paths and scripts (that are referred to dorks) so that the attackers can index them by using search engine and web crawler. A **client honeypot** is used to investigate the client-side intrusion. This type of honeypot can actively initiate requests to servers and investigate the malicious program on the server side, such as Ghost [30].

*5) **Physicality / Virtuality** (P/V):* it denotes the state of decoys as they actually exist, which can be divided into two categories: physical and virtual. A **physical honeypot** refers to a genuine computer system running on a physical machine and acting as a decoy. Indeed, physical honeypot often implies high-interaction but could have higher performance than virtual HIH. However, it is infeasible to deploy physical honeypots for each IP address in a large address space. The contrary concept is **virtual honeypot** that uses virtual decoys that need the host machine to respond to the network traffic sent to the virtual decoys [31]. We can have multiple virtual honeypots hosted concurrently by one physical machine.

Though according to the definition of honeypot, any type of information system resource can be deployed as decoy, the use of virtualization technologies has important advantages in terms of ease of management and maintenance. On one hand, all the LIHs and MIHs are virtual honeypots according to their nature of design. The decoys of LIHs are software artifice, which emulate the fingerprints of operating systems and services. On the other hand, the HIHs can be virtualized by using virtualization technologies. Galan et al. [32] summarised the virtualization technology evolution through three categories: baseline virtualization, testbed oriented virtualization and datacenter oriented virtualization. Figure 4 shows the development of the virtualization technologies used in deploying HIHs in the last 17 years (dates are approximate).

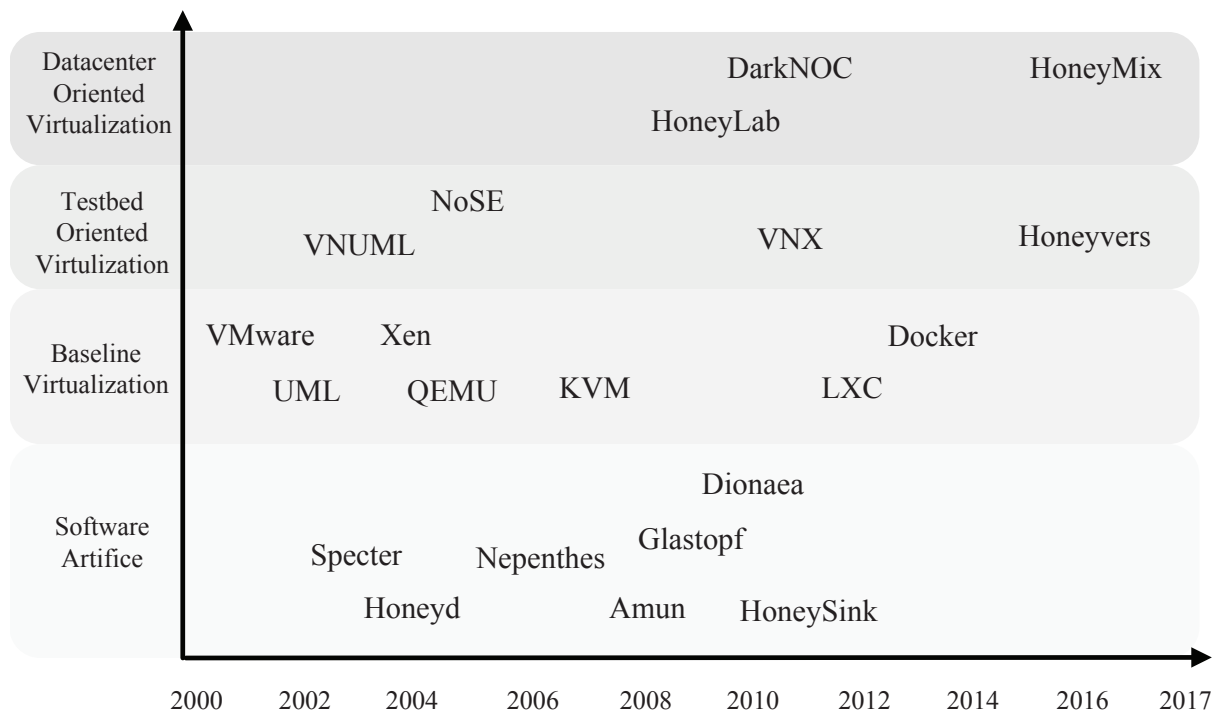

Fig. 4. Virtualization technologies for decoy deployment

For the HIHs using the baseline virtualization technologies, the first example is that the User-Mode Linux (UML) was used as the virtualization engine to mimic the Gen II Honeynet in [33], where the host machine runs the Honeywall to contain and monitor the entire virtual honeynet. The host can apply the built-in tty logging mechanism to capture the keystroke of the honeypots silently. However, the UML only enables Linux kernel-based virtual machines to run as an application within a normal Linux host. Instead, Abbasi and Harris [34] used VMware server to deploy virtual Gen III Honeynet, which can support various operation systems based on x86 architecture. Different from the previous work, it applied a multi-system virtual honeynet architecture that installs the Honeywall on a separate singe virtual machine instead of the host. It also used the Sebek to perform system activity capture in the virtual honeypots. Similarly, the KVM (Kernel-based Virtual Machine) hypervisor can also provide the emulation for different operation systems. Capalik's system [35] used the low-level surveillance of the KVM hypervisor to stealthily monitor the system activity from outside of the virtual machine, which results in the attacks having no way to bypass the surveillance. Recently, some novel lightweight virtualization technologies provide alternatives to the hypervisor-based virtualization for honeypot deployment. For example, Memari et al. [36] created virtual honeynet based on the LXC (Linux Containers), which can simultaneously create multiple Linux user-space instances through partitioning the resource of the host. The LXC based virtual machine can startup very fast but it can only emulate Linux over Linux. Another case in point is the honeypot [37] created using Docker, which can implements a high-level API to provide lightweight containers that run processes in isolation. The Docker container is even more lightweight than the LXC container since it implements the application virtualization rather than a full system virtualization.

Because it is a complex task to manually generate all the low level details for the creation of medium-to-big size honeypot scenarios, several testbed oriented virtualization technologies were proposed, which can be used to deploy virtual HIHs. VNUML (Virtual Network User Mode Linux) [38] proposed a high-level description for virtual honeynet and developed a tool to process that description automatically avoiding the user to deal with the complex low-level details. However, it only focuses on using the UML as its underlying technology so it still can only emulate Linux kernel-based virtual machines. Furthermore, some generic tools that integrate multiple virtual machine hypervisors were also proposed. NoSE (Network Simulation Environment) [39] addressed the multi-hypervisor issue through integrating a variety of virtual machine hypervisors, such as UML, Xen and QEMU, into one generic platform. The drawback of NoSE and the previous proposals is that they lack the capability of dynamic configuration for the honeynet deployment. VNX [40] is a more powerful generic virtualized tool, which integrates more hypervisors, such as UML, QEMU, KVM, LXC etc, and can even undertake dynamic configuration. Following the idea of VNX, Fan et al. proposed the Honeyvers [41] framework aimed to create and manage heterogeneous honeypots.

Apart from the tools described above, some other multi-tenant datacenter oriented virtualization technologies for HIHs deployment were also proposed. Honeylab [42] provides a platform to share IP address space and computing resources. It is an overlay of distributed infrastructure which allows security researchers to create their own desired honeypot systems without setting up distributed sensors in various geographical locations. DarkNOC [43] is designed to collect interesting traffic from different information sources, e.g. NetFlow, Snort, and Nepenthes, to analyze the data and present it to users in an efficient manner. In addition, Han et al. proposed the HoneyMix [44] system that treats honeypot as a network security function and instantiates honeypots by using Network Function Virtualization (NFV) technique.

*6) **Deployment Strategy**:* it presents the pertinence tactics of deploying the decoy(s). There are five common decoy deployment strategies: sacrificial lamb, deception ports, proximity decoys, minefield and redirection shield.

**Sacrificial lamb** is often a normal system without connections to production networks, and waits to be compromised by attacks, e.g. Argos [8] and Cuckoo Sandbox [20]. It can be a commercial off-the-shelf (COTS) computer, a router, or a switch etc. The typical implementation involves loading the operating system, configuring some applications and then leaving it on the network to see what happens. Sacrificial lambs provide a mean to analyze a compromised system down to the last byte with no possible variation. The analysis often requires numerous third-party tools. They also do not provide integrated traffic containment facilities, so will require additional network considerations.

**Deception ports** on production systems indicate simulated services disguised as well-known services. These are basically LIHs or MIHs, such as Specter [28] and Dionaea [18], which mimic various services on different ports of the system, e.g. HTTP is mimicked on port 80. These honeypots first "observe" the operating system they reside on and then portray these services according to that. The basic idea is deception so that the adversaries are "stuck-up" in solving the deception and they can be knocked down from the network.

**Proximity decoys** indicate that the decoys are deployed on the same network as production systems and possibly clone the configuration of the production systems. There will be no legal hassles to monitor the decoys. Because they are part of the same subnet where the production servers are included, and it is allowed to monitor any activity pertaining to your own network. Also, once they are in proximity to other production systems, you have ease in either re-routing traffic to the honeypots when some malicious attack is detected on the production systems, or trapping that attack. Honeyd [17] can use the free IP addresses of a certain production network to deploy and integrate the decoys into the production network, which exactly follows this deployment strategy.

**Minefield** means deploying a relatively large number of honeypots at the perimeter or the forefront of the protected network to act as landmines that "explode upon contact", by which we mean switch on the data capture function upon contact. Any scans or vulnerability detectors can exploit the contents of honeypots, sparing the production servers. So this deployment strategy can be used to capture a large amount of data. As stated, IDSs are placed at the perimeter, where they can use the contents of honeypots to reduce the probability of generating false alarms. Sinkholes, e.g. network telescopes [21], often uses this deployment strategy.

**Redirection shield** uses port redirection or traffic re-routing to forward the malicious data to the honeypots. This strategy needs the intrusion detection technology to evaluate the network traffic. If the traffic is interesting, it will be redirected into the honeypot shield to avoid the production system being attacked. Also, the shield and the production network should be coupled, tightly or loosely. The honeypots can reside in the same address space of the production network or reside on another subnet alongside the production network or even remotely. For example, Shadow Honeypots [5] following this deployment strategy use the shadow application as a shield dealing with the malicious traffic for anomaly-based detection.

*7)* ***Resource Type:*** it denotes the type of information system resource available for the attacks. Most honeypots provide or emulate general attacked resources, which are aimed at detection of more than one attack technique. Currently, many specific attacked resource oriented honeypot systems have been proposed, which can be defined as follows:

**Web application honeypots** are tools aimed at detection of attacks on web application, e.g. Glastopf [29];

**VoIP honeypots** are used to capture threats in internet telephony (Voice over IP), e.g. Artemisa [45];

**SSH honeypots** are oriented on secure shell (SSH) attacks, e.g. Cowrie [19];

**Bluetooth honeypots** are aimed to capture the attacks propagating through the Bluetooth device, such as the Bluepot [46];

**USB honeypots** are used to investigate arbitrary malware on USB storage devices, e.g. Ghost USB Honeypot [30];

**SCADA/ICS honeypots** emulate industrial control system resources, e.g. Conpot [47];

**IoT honeypots** are used to capture the attacks that target IoT devices, such as the IoTPOT [48];

**IPv6 network honeypots** are tools used to capture attacks targeting IPv6 networks, e.g. Hyhoneydv6 [49].

### B. *Functions of security program*

As previously stated, the security program aims to carry out all the security related functions, such as attack monitoring, prevention, detection, response and profiling. This subsection describes all these function in detail.

*1)* ***Attack Monitoring:*** it is aimed to log all the intrusion events and malicious behaviors to allow a further investigation. Two critical layers of data can be identified: **network activity** (every inbound and outbound, connection, packet and its header information as well as its payload, etc.), and **system activity** (keystroke, system call, rootkits, etc.).

Surveying the techniques for capturing and collecting network data, particularly in the case of cooperative honeypots from distributed decoys, two widely used network data forwarding methods are found: tunneling and application-level proxying. **Tunneling** is used when some distributed decoys, such as network telescopes, darknet, and blackholes, are placed in a different location where the processing backends are. As the decoys are assigned a portion of routed IP address space corresponding to its physical location, a tunnel mechanism based on a tunneling protocol like GRE has to be used to transport data packets to the backends. By using tunnels, the decoy backends seem to be directly deployed in the production network, being the tunnel almost invisible to "traceroute", although the tunnel will add some latency and modify the MTU. Some hybrid systems [26], [50] uses GRE tunnel to forward the inbound data from the frontends to the backends. **Application-level proxying** consist of transporting the content of the packets to the backends by means of application specific proxies. The application-level proxies are also known as application-level gateways, and are available for common Internet services, e.g. an HTTP proxy is used for Web access and an FTP proxy is used for file transfers. Honeyd [17] provides an application-level proxying functionality. For instance, on TCP port 23, Honeyd can be configured to automatically proxy traffic to another machine's Telnet port.

In contrast, the generic so called "circuit-level" proxies (that conceptually work at the session layer of the OSI model) give support to multiple applications. For example, SOCKS is and IP-based circuit-level proxy server that supports applications using TCP and UDP. Application-level proxies provide better support for the additional capabilities of each protocol than circuit-level ones do (e.g. application-level proxies can better support virus scanning). Also, they are client-neutral and require no special software components or operating system on the client computer to enable the client to communicate with servers through the proxy.

On the other hand, system activity monitoring needs capturing the malicious activity in HIHs. According to the requirement of stealthy capturing, some approaches have been proposed. Sebek [51] and Qebek [52] are examples of the first monitoring tools used to capture the system activity in honeypots. They modify the system kernel by adding new kernel modules that capture system activity in a supposedly hard to detect way. However, there are nowadays some techniques can detect the presence of this type of kernel modules installed inside the honeypots. The well-known CWSandbox [53] uses the in-line code overwriting approach to hook the API function in order to observe the malware behavior without being noticed. However, this approach still has the possibility of being detected.

In order to address this drawback, Jiang and Wang [54] proposed another monitoring approach called **"out-of-the-box"**, which uses the virtualization hypervisor to monitor the activity in guest virtual machines (VM). VMI-Honeymon [13] uses a Volatility extension to call the API of the Xen Access successor LibVMI to access the memory of the guest VM. LibVMI [55] is a C library with Python bindings based virtual machine introspection which can support a variety of virtual machine hypervisors, such as Xen, KVM, etc. It is easy to monitor the low-level details of a running virtual honeypot by viewing its memory trapping on hardware events and accessing the vCPU registers. There are some other virtual machine introspection based approaches that can analyze malware and meanwhile make it harder for the malware to detect them, such as Livewire [56], VMScope [54], Lares [57], VMWatcher [58], etc. However, only some of them are open-source projects and still maintained, which are listed in Table I that compares their supported hypervisors and operating systems.

TABLE I
VIRTUAL HONEYPOT INTROSPECTION SOLUTIONS

| Solution | VM Hypervisor | Supported OS |
|---|---|---|
| Argos [8] | QEMU | Windows |
| Nitro [59] | QEMU | Windows |
| Timescope [60] | QEMU | Linux |
| Virtuoso [61] | QEMU | Windows, Linux, OS X, Haiku |
| DRAKVUF [62] | Xen | Windows |
| Cuckoo [20] | KVM | Windows, Linux, OS X, Android |

*2) **Attack Prevention:*** it is aimed to deter or block intrusion. This function can be carried out by several approaches: data filtering, tarpitting and containment.

**Filtering** consists of discarding the data traffic which is typically specified by means of filtering rules. Basically, there are two filtering mechanisms: source-destination based and content based. On one hand, the **source-destination based filtering mechanism** examines the header information (mainly source and destination addresses and ports and protocol) of each packet to make the discarding decision. This mechanism is effective at reducing the amount of repeated traffic into a non-redundant manageable data. iSinks [25] uses a filtering strategy consisting of analyzing the connections established with the first N destination IPs per every source IP. Pang et al. [63] improved the filtering mechanisms taking into account, for example, the source port, destination and connection. Bailey et al. [64] improved the source-destination based filtering mechanism through expanding the individual darknets into multiple darknets for observing the global behavior and the source distribution. On the other hand, **content based filtering mechanism** consist of inspecting the content or payload of the packets to make discarding decision. Bailey et al. [10] proposed content prevalence as a filtering mechanism by inspecting the first packet including new payload. Content prevalence analyzes the distribution of content sequences in payloads, and can generate alert when a specific piece of content sequence becomes widely prevalent. Similarly, IMS [24] proposed a caching mechanism to avoid record duplicated payload through recording the first payload packets in order to reduce disk utilization. A potential drawback of packet inspection based filtering mechanism is that it will be unable to make decision until the session has been established and at least the first packet of content or payload has been received. Also, SweetBait [65] uses whitelists to filter the traffic matching benevolent patterns to conduct zero-day worm detection. RolePlayer [66] can emulate both the client and the server side of an application session in order to replay and filter variant well-known attacks. Shadow honeypot [5] uses signature-based IDS to filter the well-known attacks and then applies anomaly-based IDS to filter the input into suspect traffic for further investigation.

**Tarpitting** consists of purposely slow down the attacking progress, worm propagation, virus sprawl, etc. Collapsar's [67] tarpit module restricts the outgoing attack from the honeypot by throttling the packet rate sent. Honeywall [68] is also a tarpit device that can limit the number of outgoing connections. It can block any outbound connections in the case it is capturing automated attacks, or when investigating manual attacks, it can be programmed to allow some outbound connections, such as 5 to 10 connections per hour. However, the brute strict data tarpitting will raise the adversary's suspicion, as well as bringing the chance of being detected, which will lead to impede data capture.

**Containment** is another approach to prevent the adversary to use the compromised honeypots to attack other non-Honeypot systems, through confining the attack in the honeypot environment. In order to reduce the risk of being detected, it redirects the outbound attacks back to other honeypots, rather than limiting the number of outgoing connections or discarding them. Alata et al. [69] implemented such outgoing connection redirection mechanism through modifying the Linux system kernel. Whereas, the outgoing traffic redirection has some drawback as well: it uses the "in-the-box" approach, which allows some advanced adversaries to detect the redirection module.

*3) **Attack Detection**:* this function aims to detect intrusion and generate alerts. Commonly, there are two detection approaches: signature-based and anomaly-based.

**Signature-based detection** is based on detecting the well-known attacks by recognizing malicious patterns. This approach is often used in production environments to discover unauthorised activity and generate alerts to the administrator. Unlike the production traffic captured by IDSs, the traffic received by honeypots will almost always correspond to malicious activities, as the honeypots have no production value. Attack detection honeypots therefore have a highly reduced false alarm rate. This type of honeypot is often called **production honeypot** and is aimed to emulate well-known vulnerabilities to lure intrusion so that intruders will be deceived by being forced to waste time to interact with the honeypots. Production honeypots are often LIHs and MIHs that have little or no interaction with the attacker in order to minimize the risk of infecting other production systems in the context. Besides, the performance and response time should be guaranteed for production ones. For example, the production honeypot Dionaea [18] can simulate multiple well-known services to carry out signature-based detection.

On the other hand, **anomaly-based detection** means detecting unknown attacks by discovering deviations from normal behaviour patterns. The honeypots using this detection approach are always used in the research environment as research honeypots. A **research honeypot** is designed to detect anomaly attacks and investigate the unknown signatures. Thus, research honeypots are often more powerful than production honeypots. HIHs and hybrid honeypot systems are always used as research ones to provide full functional systems. A wider assortment of data can be captured to facilitate further investigation for many purposes. Also, research honeypots are a step ahead than production ones. The signature of new attack generated by research ones are often used to serve the production ones for attack detection. This can lead to early warning and prediction of future attacks and exploits. At present, a number of anomaly-based detection techniques have been proposed in the context of honeypot research. For example, Argos [8] applies the dynamic taint analysis [70] to detect zero-day attacks and meanwhile generate the new signatures, Honeycomb [71] uses the longest common substring (LCS) algorithm to detect repeating patterns in order to spot worms, and Bailey's system[10] performs system behavior profiling by comparing the infected virtual filesystem to the uninfected one. In addition, some current learning techniques, such as the deep learning approach for NIDS [72], can also be used in decoy so as to acquire new detection skills for identify unknown attacks.

Apart from the traditional IDS techniques, Sekar et al. [73] proposed the Specification-based Anomaly Detection approach using the supervised method to develop specification, instead of using unsupervised machine learning techniques, in order to identify legitimate behaviors and detect unknown attacks as deviation from a norm. It improves the effectiveness of anomaly detection that lacks in signature-based approaches, but also minimizes the large number of false positive produced by anomaly-based techniques.

*4) **Attack Response**:* it relates to the measures taken to respond to attacks or adapt to intrusion events in terms of certain requirements. Basically, honeypots can take two type of reactions: traffic redirection and decoy reconfiguration.

*a) **Traffic redirection**:* it is used to control how the traffic is sent to the appropriate destination. For example, hybrid honeypots redirects the malicious traffic from LIH to isolated HIH for further investigation. We mainly take into account two redirection techniques: Flow-based routing and TCP connection replaying.

**Flow-based routing** is the routing technique where packets are routed from source to destination, based on selecting the path that satisfies some requirements such as QoS, load balance, security, etc. This mechanism is based on the same principles used for normal routing in networks, but it is applied to more specific data flows. Kohler et al. proposed the flexible and configurable Click modular router, which is made of simple packet processing modules which are combined in a service chain in order to build complex and efficient network services that can be used in this case to do flow based routing [74]. There are several cooperative honeypot systems using Click framework to facilitate the data control. For example, the Potemkin gateway router and the GQ gateway are based on the Click modular router. In addition, with the rapid growth of software-defined networking (SDN), OpenFlow was designed to allow users to programmatically control real switches (from companies like Cisco, HP, etc.) by means of applications running on SDN controller frameworks. The SDN controller can facilitate a fine-grained dynamic control of traffic by means of the flow table entries configured on each OpenFlow based switches. In the near future, the programmable SDN based network architectures will increasingly take the role of the data control for honeypot systems [75].

**TCP connection replaying** is a connection handover technique aimed to seamlessly transfer one TCP socket endpoint from one node to another node. When an interesting connection is established between the attacker and the LIH, a TCP connection handoff mechanism is needed to redirect the connection from the LIH to a HIH for further investigation. It transfers the established TCP state of the socket endpoint from the original node to the new one, and then the new node can continue the conversation with the other TCP endpoint directly. Bailey's system [10] uses a connection handoff mechanism for traffic redirection. In order to avoid conserving the state of every connection, the connection handoff mechanism makes the redirection decision based on the first payload packet of each connection. However, the author did not unveil the technical detail about the connection handoff. Similarly, Honeybrid gateway [76] uses the connection replay mechanism to implement traffic transparent redirection between LIHs and HIHs. Nevertheless,

Honeybrid revealed the technical details of the gateway, which is a TCP replay proxy using libnetfilter_queue [77] to process packets. The connection handoff mechanism based on TCP replay is able to provide stealthy redirection for automated malwares. In [78], Lin et al. proposed a transparent and secure network environment which can allow the automated malwares to attack or propagate, but under a stealthy control. Although the TCP/IP stateful traffic replay can facilitate transparent TCP connection handoff, it cannot solve the identical-fingerprint problem, which means the LIH and HIH have different fingerprints (e.g. IP and MAC addresses). This problem leaves the opportunity to the skilled adversaries to detect the honeypot environment. VMI-Honeymon [79] provided a novel solution that retains the MAC and IP address of the original HIH for cloned HIHs but creates separate network bridges to isolate them so as to avoid address collisions. Most recently, Fan and Fernández [80] proposed a novel SDN based stealthy TCP connection handover mechanism that solved this problem through using different ports of OpenFlow based switch to isolate the honeypots with identical-fingerprint.

*b) **Decoy reconfiguration**:* it is designed to timely adapt decoy's state to specific events, which could be intrusion events, state variation of objective targets, etc. As stated, static honeypot system lacks the capability to reconfigure the decoy in time. That is a critical disadvantage in the complex and dynamic network scenarios where the honeypots are deployed. Several approaches have been proposed to address this problem, which can be roughly categorised into dynamic cloning and dynamic catering.

**Dynamic cloning** is aimed to synchronously emulate the real production targets including network topology, operating system fingerprints, services, open ports, etc. It is designed to rapidly revolutionize the configuration and deployment by monitoring and learning the target organization networks in real time. Thus, the dynamic cloning has two phases. The first phase is called network discovery which is used to collect the information of target network. The second phase is called honeypot deployment which refers to deploy decoys emulating the target systems. There are two ways to discover the targets: passive and active fingerprinting. Hecker et al. [81] discussed both the two ways for the network discovery and automated honeynet cloning. Passive fingerprinting tools, such as p0f [82], can sniff the traffic, determine active systems and open ports in the target scenario, and meanwhile make little traffic noise. However, the main problem of this approach is that it does not discover the systems that do not generate any production traffic. Instead, active probing tools, such as Nmap [83], can discover all open ports on the objective system, even if there is no production traffic to those ports, at the price of generating some extra production traffic. In [9], a dynamic hybrid honeypot systems is proposed for intrusion detection. It consists of a combination of LIHs and HIHs, and relies on active probing to get information of the organization networks for emulation. In the network discovery phase, the active probing tool Nmap [83] is used to determine the active systems and open ports. Then in the honeypot deployment phase, LIHs are created periodically by Honeyd [17] to represent the production systems, and it also used virtual HIHs to receive the redirected traffic from LIHs, but the dynamic deployment of HIHs was not mentioned.

**Dynamic catering** is used to create catering honeypots for certain attacks, gradually escalating interaction level for malicious data capture, redeploying the honeypots when intrusion activity is detected. It follows the idea of creating and deploying the honeypots on demand timely to increase data capture efficiency. Potemkin [26] used dynamically created HIHs on physical servers to achieve efficient resources usage. It employs a network gateway, to which routers all over the Internet are configured to tunnel an address prefix, as an agent to take responsibility of sending traffic to a honeyfarm server. The gateway instructs the virtual machine monitor (VMM) that runs on each physical server to create a new HIH on demand for each active destination IP address. Otherwise, if one HIH is idle, the VMM will destroy it and reclaim the resources when being instructed by the gateway. Similarly, VMI-Honeymon [79] clone VMs through restoring the memory snapshot with configuration on a QEMU copy-on-write (qcow2) filesystem. The newly created virtual HIH runs the system and applications in exactly the same fingerprints with the cloned one to investigate the attacks.

*5) **Attack Profiling**:* it is the extrapolation of attack information in order to analyze malicious activity, as well as unveiling the intrusion motives. McGrew and Vaughn, Jr. [84] ever indicated that an attack profile should contain these following attributes:

**Motivation** describes the reason of the attack;

**Breadth/Depth** presents the scope of the attack and the degree of the impact to the attacked system;

**Sophistication** shows the level of technical expertise to carry out the attack;

**Concealment** describes the measures used for hiding the evidence of the attack;

**Attacker(s)** defines the role behind the attack: individual or a group of adversaries, or at least identifies the source of the attack, e.g. automated malware;

**Vulnerability** is the flaw that can be exploited by the attack;

**Tools** are the software used to carry out attacks, including shellcodes, back-doors, rootkits, and other software uploaded to the system to perform the rest of the attack.

Among these attributes, some can be achieved by directly applying the captured honeypot data. Through statistically analyzing the log information, including the attack source, destination and frequency, as well as the infection degree on the HIH, we can identify the breadth and the depth of the attack separately. Also, the concealment and tools can be displayed through observing the adversary's activity on the honeypot. We call this approach using basic statistics on the log information is **direct information based** attack profiling. Some honeypots, e.g. Honeyd [17] and Dionaea [18], do use the source, destination and timestamp of an attack based on IP information to described the attack profile.

However, the other attributes have to be revealed by further derived information. Motivation can be inferred according to the insights into the activity on HIH. Identifying the attacker and the sophistication needs in-depth observation and forensics on the interaction between the attacker and the honeypot. The unknown vulnerabilities often need advanced detection techniques. Therefore, **derived information based** attack profiling is much more complex, since it tries to access and explain the fundamental cause of the attack. For this purpose, basic statistics do not suffice any more, it is necessary to apply interdisciplinary approaches, e.g. association rule mining, neuronal networks, virtual machine introspection, etc. Currently, plenty of techniques have been presented to analyze the malicious data: Nawrocki et al. [11] reviewed the approaches for honeypot data analysis; Egele et al. [85] surveyed the automated dynamic malware analysis techniques and tools; Rieck et al. [86] presented the research on honeypot system behavior analysis through machine learning.

### *C. The design space and constraints*

Depending on the classification scheme, the honeypot designer can theoretically gain at most 103680 different combinations of classes, which can provide a global view of design space of homogeneous honeypots. However, we have to note that there are several mutual restrictions between some features (see Fig. 5) that can lead to the design space shrink.

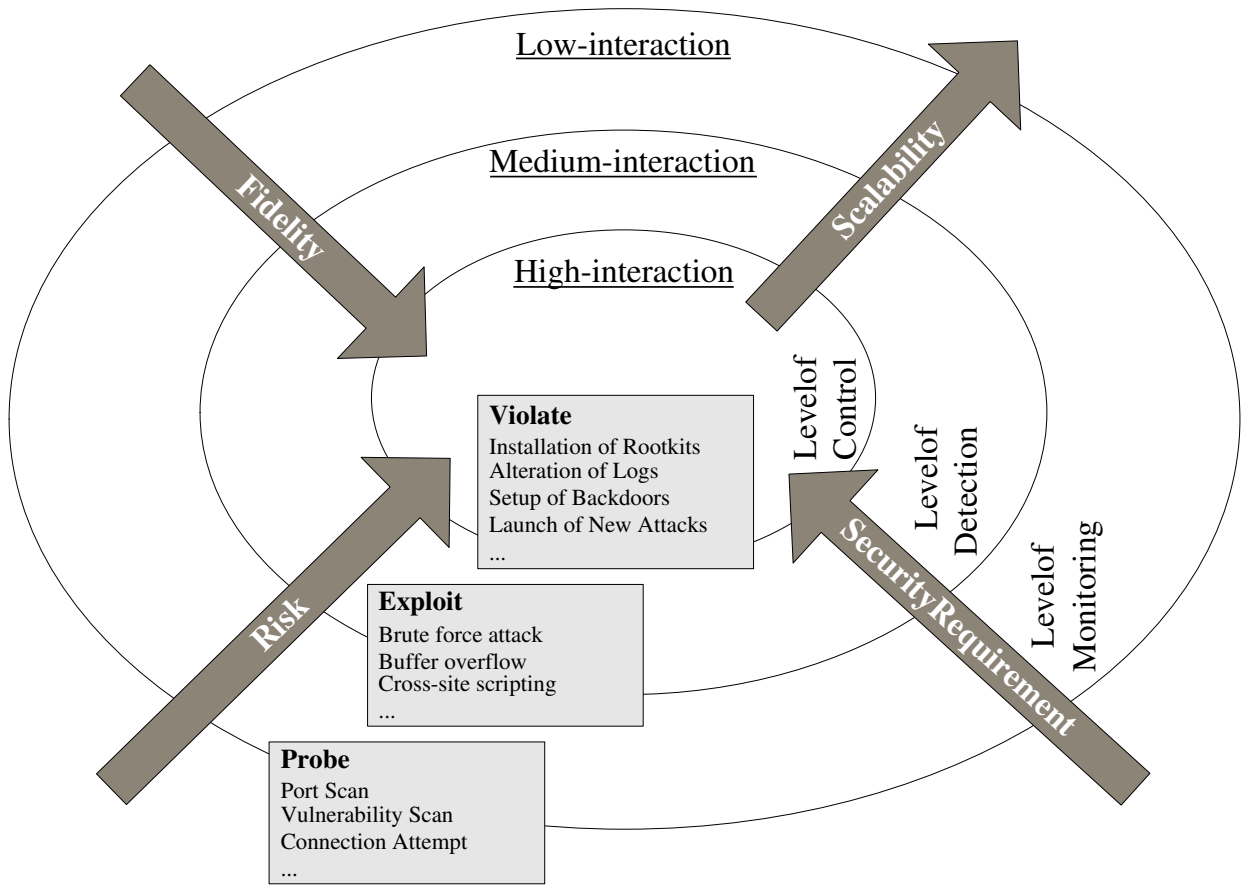

Fig. 5. Constraints between main features

From the point of view of resisting attacks, malicious data can be captured through honeypots being probed, attacked and compromised, which are exactly corresponding to the three phases of cyber attack, i.e. probe, exploit and violate. Attacks always begin at probing large-scale IP networks in order to find vulnerable objectives, and then they exploit the vulnerabilities to compromise the objectives, and finally, if the compromised systems are worth further utilizing, the adversary will violate them, e.g. install rootkits, setup backdoors, and launch new attacks etc. The large-scale probing will produce high network traffic load but it will not translate into a useful system activity. Then a part of the probed systems which have vulnerabilities will be attacked. So, in the exploiting phase, the scale of attacking objectives is reduced, but the data quality is enhanced, i.e. attacking traffic will include malicious payload. In the violating phase, only a small part of compromised systems or even several specific objectives will be involved, and the data quality become very high because any unauthorized system data is worth recording for further investigation. Therefore, every phase produces a different data quantity and quality. The fidelity and scalability features are highly related with the three attack phases.

However, fidelity and scalability is a pair of mutual-condition features in a certain type of decoy. On one hand, for approaching the task of capturing high quality data, decoy has to escalate the interaction level, however a higher interaction level will cause a higher risk of being compromised, so that the honeypot have to enhance the security program to protect the decoy. On the other hand, higher interaction guarantees the fidelity but sacrifices the scalability, which means it will result in failing to capture adequate data from large-scale IP network spaces. So, it needs to seek a good balance between those two features in order to optimize the use of honeypot resources. The cooperative honeypots, particularly, the hybrid honeypots are developed to overcome those issues.

Also, according to the above discussion, we can note that the attack profiling of security program is highly related to the fidelity of decoy. If the security program wants to perform attack profiling based on derived information, the decoy needs high interaction level to derive deep enough information (i.e. the system activity) about the attack. Otherwise, if the honeypot is a LIH or MIH, the attack profiling can only use the direct information.

Furthermore, the adaptability is highly relies on the P/V. It is observable that physical honeypots are often static, while the virtualization technology has made it easy to create dynamic honeypots. The software artifice is the easiest way to carry out dynamic configuration, and at present, the virtual machine based HIHs are increasingly convenient to perform dynamic reconfiguration. So, the decoy reconfiguration of attack response is also tightly related to the P/V.

Overall, the design space can help predicting the future honeypot design in theory, while the constraints among different features can provide a more practical view for the designer to implement honeypots under specific technical environment.

## IV. Honeypot Development Analysis

This section surveys a number of typical honeypots, including the independent and cooperative ones, by applying the D-P based taxonomy. The comparison of these selected honeypots is illustrated in Table II. It shows the proposed D-P based taxonomy can fully classify those different honeypots. Also, there are some development tendency can be discovered that will be analyzed in the following subsections.

### *A. Hybrid honeypots*

According to the requirement of decoupling and achieving the optimization of both the fidelity and scalability, many cooperative honeypots (particularly, the hybrid honeypots) have been developed ( see the right part of Table II). Commonly, a typical hybrid honeypot consists of three subsystems: frontends, controller and backends. The backends can be HIHs or honeyfarm, the frontends can be LIHs (or MIHs) or sinkholes for monitoring large-scale routed

TABLE II
COMPARISON OF HONEYPOT SYSTEMS IN TERMS OF D-P BASED TAXONOMY

| | **Independent Honeypots** | | | | | | | | | | | | **Cooperative Honeypots** | | | | | | | | | | | | | |
|---|---|---|---|---|---|---|---|---|---|---|---|---|---|---|---|---|---|---|---|---|---|---|---|---|---|---|
| | Specter(2003)[28] | Network Telescopes(2004)[21] | Honeyd(2004)[17] | Argos(2006)[8] | Glastopf(2009)[29] | Dionaea(2011)[18] | Artemisa(2011)[45] | Bluepot(2012)[46] | Ghost(2012)[30] | Conpot(2013)[47] | Cuckoo Sandbox(2014)[20] | Cowrie(2015)[19] | Bailey's System(2004)[10] | Collapsar(2004)[67] | Shadow Honeypots(2005)[5] | Potemkin(2005)[26] | Gen II Honeynet(2005)[87] | Artail's System(2006)[9] | GQ(2006)[50] | SweetBait(2007)[65] | Honeybrid(2008)[76] | SGNET(2008)[88] | Li's System(2009)[89] | VMI-Honeymon(2013)[79] | IoTPOT(2015)[48] | Hyhoneydv6(2015)[49] |
| **Fidelity** | | | | | | | | | | | | | | | | | | | | | | | | | | |
| Low | ● | ● | ● | ○ | ● | ○ | ● | ○ | ○ | ● | ○ | ○ | ● | ○ | ○ | ○ | ○ | ● | ● | ● | ● | ● | ● | ○ | ● | ● |
| Medium | ○ | ○ | ○ | ○ | ○ | ● | ○ | ● | ● | ○ | ○ | ● | ○ | ○ | ○ | ○ | ○ | ○ | ○ | ○ | ○ | ○ | ○ | ● | ○ | ○ |
| High | ○ | ○ | ○ | ● | ○ | ○ | ○ | ○ | ○ | ○ | ● | ○ | ● | ● | ● | ● | ● | ● | ● | ● | ● | ● | ● | ● | ● | ● |
| **Physicality / Virtuality** (P/V) | | | | | | | | | | | | | | | | | | | | | | | | | | |
| Physical | ○ | ○ | ○ | ○ | ○ | ○ | ○ | ○ | ○ | ○ | ○ | ○ | ○ | ○ | ● | ○ | ● | ● | ○ | ○ | ○ | ○ | ● | ○ | ○ | ○ |
| Virtual | | | | | | | | | | | | | | | | | | | | | | | | | | |
| Software artifice | ● | ● | ● | ○ | ● | ● | ● | ● | ● | ● | ○ | ● | ● | ○ | ● | ○ | ○ | ● | ● | ● | ● | ● | ● | ● | ● | ● |
| VMware | ○ | ○ | ○ | ○ | ○ | ○ | ○ | ○ | ○ | ○ | ○ | ○ | ● | ○ | ○ | ○ | ● | ○ | ● | ○ | ● | ○ | ● | ○ | ○ | ○ |
| UML | ○ | ○ | ○ | ○ | ○ | ○ | ○ | ○ | ○ | ○ | ○ | ○ | ○ | ● | ○ | ○ | ● | ○ | ○ | ○ | ○ | ○ | ○ | ○ | ○ | ○ |
| Xen | ○ | ○ | ○ | ○ | ○ | ○ | ○ | ○ | ○ | ○ | ○ | ○ | ○ | ○ | ○ | ● | ○ | ○ | ○ | ○ | ○ | ○ | ○ | ● | ○ | ○ |
| QEMU | ○ | ○ | ○ | ● | ○ | ○ | ○ | ○ | ○ | ○ | ○ | ○ | ○ | ○ | ○ | ○ | ○ | ○ | ○ | ● | ○ | ● | ○ | ○ | ● | ● |
| KVM | ○ | ○ | ○ | ○ | ○ | ○ | ○ | ○ | ○ | ○ | ● | ○ | ○ | ○ | ○ | ○ | ○ | ○ | ○ | ○ | ○ | ○ | ○ | ○ | ○ | ○ |
| **Scalability** | | | | | | | | | | | | | | | | | | | | | | | | | | |
| Unscalable | ○ | ○ | ○ | ● | ○ | ● | ○ | ○ | ● | ○ | ● | ○ | ○ | ○ | ○ | ○ | ○ | ○ | ○ | ○ | ○ | ○ | ○ | ○ | ○ | ○ |
| Scalable | ● | ● | ● | ○ | ● | ○ | ● | ● | ○ | ● | ○ | ● | ● | ● | ● | ● | ● | ● | ● | ● | ● | ● | ● | ● | ● | ● |
| **Adaptability** | | | | | | | | | | | | | | | | | | | | | | | | | | |
| Static | ● | ● | ○ | ● | ○ | ● | ○ | ● | ● | ● | ● | ● | ○ | ○ | ○ | ○ | ● | ○ | ○ | ○ | ○ | ○ | ○ | ○ | ○ | ○ |
| Dynamic | ○ | ○ | ● | ○ | ● | ○ | ● | ○ | ○ | ○ | ○ | ○ | ● | ● | ● | ● | ○ | ● | ● | ● | ● | ● | ● | ● | ● | ● |
| **Role** | | | | | | | | | | | | | | | | | | | | | | | | | | |
| Server | ● | ● | ● | ● | ● | ● | ● | ○ | ○ | ● | ○ | ● | ● | ● | ● | ● | ● | ● | ● | ● | ● | ● | ● | ● | ● | ● |
| Client | ○ | ○ | ○ | ○ | ○ | ○ | ○ | ● | ● | ○ | ● | ○ | ○ | ● | ● | ○ | ○ | ○ | ○ | ○ | ○ | ○ | ● | ○ | ○ | ○ |
| **Deployment Strategy** | | | | | | | | | | | | | | | | | | | | | | | | | | |
| Sacrificial lamb | ○ | ○ | ○ | ● | ○ | ○ | ○ | ○ | ● | ○ | ● | ○ | ● | ● | ○ | ● | ● | ● | ● | ● | ● | ● | ○ | ● | ● | ● |
| Deception ports | ● | ○ | ○ | ○ | ● | ● | ○ | ● | ○ | ● | ○ | ● | ○ | ○ | ○ | ○ | ○ | ○ | ○ | ○ | ○ | ○ | ○ | ● | ● | ○ |
| Proximity decoys | ○ | ○ | ● | ○ | ○ | ○ | ● | ○ | ○ | ○ | ○ | ○ | ● | ○ | ○ | ○ | ● | ● | ○ | ● | ● | ● | ● | ○ | ○ | ● |
| Minefield | ○ | ● | ○ | ○ | ○ | ○ | ○ | ○ | ○ | ○ | ○ | ○ | ○ | ● | ○ | ● | ○ | ○ | ● | ○ | ○ | ○ | ○ | ○ | ○ | ○ |
| Redirection shield | ○ | ○ | ○ | ○ | ○ | ○ | ○ | ○ | ○ | ○ | ○ | ○ | ○ | ○ | ● | ○ | ○ | ○ | ○ | ○ | ○ | ○ | ○ | ○ | ○ | ○ |
| **Resource Type** | | | | | | | | | | | | | | | | | | | | | | | | | | |
| General | ● | ● | ● | ● | ○ | ● | ○ | ○ | ○ | ○ | ● | ○ | ● | ● | ● | ● | ● | ● | ● | ● | ● | ● | ● | ● | ○ | ○ |
| Web application | ○ | ○ | ○ | ○ | ● | ○ | ○ | ○ | ○ | ○ | ○ | ○ | ○ | ○ | ○ | ○ | ○ | ○ | ○ | ○ | ○ | ○ | ○ | ○ | ○ | ○ |
| VoIP | ○ | ○ | ○ | ○ | ○ | ○ | ● | ○ | ○ | ○ | ○ | ○ | ○ | ○ | ○ | ○ | ○ | ○ | ○ | ○ | ○ | ○ | ○ | ○ | ○ | ○ |
| SSH | ○ | ○ | ○ | ○ | ○ | ○ | ○ | ○ | ○ | ○ | ○ | ● | ○ | ○ | ○ | ○ | ○ | ○ | ○ | ○ | ○ | ○ | ○ | ○ | ○ | ○ |
| Bluetooth | ○ | ○ | ○ | ○ | ○ | ○ | ○ | ● | ○ | ○ | ○ | ○ | ○ | ○ | ○ | ○ | ○ | ○ | ○ | ○ | ○ | ○ | ○ | ○ | ○ | ○ |
| USB | ○ | ○ | ○ | ○ | ○ | ○ | ○ | ○ | ● | ○ | ○ | ○ | ○ | ○ | ○ | ○ | ○ | ○ | ○ | ○ | ○ | ○ | ○ | ○ | ○ | ○ |
| SCADA/ICS | ○ | ○ | ○ | ○ | ○ | ○ | ○ | ○ | ○ | ● | ○ | ○ | ○ | ○ | ○ | ○ | ○ | ○ | ○ | ○ | ○ | ○ | ○ | ○ | ○ | ○ |
| IoT | ○ | ○ | ○ | ○ | ○ | ○ | ○ | ○ | ○ | ○ | ○ | ○ | ○ | ○ | ○ | ○ | ○ | ○ | ○ | ○ | ○ | ○ | ○ | ○ | ● | ○ |
| IPv6 networks | ○ | ○ | ○ | ○ | ○ | ○ | ○ | ○ | ○ | ○ | ○ | ○ | ○ | ○ | ○ | ○ | ○ | ○ | ○ | ○ | ○ | ○ | ○ | ○ | ○ | ● |
| **Attack Monitoring** | | | | | | | | | | | | | | | | | | | | | | | | | | |
| Application layer proxying | ○ | ○ | ○ | ○ | ○ | ○ | ○ | ○ | ○ | ○ | ○ | ○ | ○ | ○ | ● | ○ | ○ | ● | ○ | ○ | ○ | ○ | ○ | ○ | ○ | ● |
| GRE tunneling | ○ | ○ | ○ | ○ | ○ | ○ | ○ | ○ | ○ | ○ | ○ | ○ | ○ | ● | ○ | ● | ○ | ○ | ● | ○ | ○ | ○ | ○ | ○ | ○ | ○ |
| In-the-box | ○ | ○ | ○ | ○ | ○ | ○ | ○ | ○ | ○ | ○ | ○ | ○ | ● | ● | ● | ○ | ● | ● | ○ | ○ | ● | ○ | ○ | ○ | ○ | ○ |
| Out-of-the-box | ○ | ○ | ○ | ● | ○ | ○ | ○ | ○ | ○ | ○ | ● | ○ | ○ | ○ | ○ | ● | ○ | ○ | ● | ● | ○ | ● | ○ | ● | ○ | ○ |
| **Attack Prevention** | | | | | | | | | | | | | | | | | | | | | | | | | | |
| Source-destination filtering | ○ | ○ | ○ | ○ | ○ | ○ | ○ | ○ | ○ | ○ | ○ | ○ | ○ | ○ | ○ | ● | ○ | ○ | ● | ○ | ● | ○ | ○ | ● | ○ | ○ |
| Content filtering | ○ | ○ | ○ | ○ | ○ | ○ | ○ | ○ | ○ | ○ | ○ | ○ | ● | ○ | ● | ● | ○ | ○ | ● | ● | ● | ○ | ○ | ● | ○ | ○ |
| Tarpitting | ○ | ○ | ○ | ○ | ○ | ○ | ○ | ○ | ○ | ○ | ○ | ○ | ○ | ● | ○ | ○ | ● | ● | ● | ○ | ● | ○ | ○ | ● | ○ | ○ |
| Containment | ○ | ○ | ○ | ○ | ○ | ○ | ○ | ○ | ○ | ○ | ○ | ○ | ● | ○ | ○ | ● | ○ | ○ | ● | ○ | ○ | ○ | ○ | ● | ○ | ○ |
| **Attack Detection** | | | | | | | | | | | | | | | | | | | | | | | | | | |
| Signature-based | ● | ○ | ○ | ○ | ● | ● | ○ | ● | ● | ● | ● | ○ | ○ | ● | ○ | ○ | ● | ● | ○ | ○ | ● | ○ | ● | ● | ● | ○ |
| Anomaly-based | ○ | ○ | ○ | ● | ○ | ○ | ○ | ○ | ● | ○ | ○ | ○ | ● | ○ | ● | ○ | ○ | ○ | ○ | ● | ● | ● | ○ | ● | ○ | ○ |
| **Attack Response** | | | | | | | | | | | | | | | | | | | | | | | | | | |
| Flow based routing | ○ | ○ | ○ | ○ | ○ | ○ | ○ | ○ | ○ | ○ | ○ | ○ | ○ | ● | ○ | ● | ○ | ○ | ○ | ○ | ○ | ○ | ○ | ○ | ○ | ○ |
| TCP connection replaying | ○ | ○ | ○ | ○ | ○ | ○ | ○ | ○ | ○ | ○ | ○ | ○ | ● | ○ | ○ | ○ | ○ | ○ | ○ | ○ | ● | ● | ○ | ● | ○ | ○ |
| Dynamic cloning | ○ | ○ | ○ | ○ | ○ | ○ | ○ | ○ | ○ | ○ | ○ | ○ | ○ | ○ | ○ | ○ | ○ | ● | ○ | ○ | ○ | ○ | ○ | ○ | ○ | ○ |
| Dynamic catering | ○ | ○ | ○ | ○ | ○ | ○ | ○ | ○ | ○ | ○ | ○ | ○ | ● | ○ | ○ | ● | ○ | ○ | ● | ○ | ○ | ○ | ○ | ● | ○ | ● |
| **Attack Profiling** | | | | | | | | | | | | | | | | | | | | | | | | | | |
| Direct information | ● | ● | ● | ○ | ● | ● | ● | ● | ● | ● | ○ | ● | ● | ● | ● | ● | ● | ● | ● | ● | ● | ● | ● | ● | ● | ● |
| Derived information | ○ | ○ | ● | ● | ○ | ○ | ○ | ○ | ● | ○ | ● | ○ | ● | ● | ● | ● | ● | ● | ● | ● | ● | ● | ○ | ● | ○ | ○ |

IP address space, and the controller can be Honeywall, Click modular router, Honeybrid gateway, etc. These three subsystems facilitate a number of functions. The frontends often provide low interaction to the attacks, because their main objective is to capture network data. However, they need to discard the uninteresting traffic in order to get fine-grained data. The controllers are used to perform the functions of data control as well as system dynamic configuration in a hidden way. The backends are aimed to perform stealthy system data capture and perform data analysis such as digital forensics to unveil the attacks' skills, tactics and motives. Table III shows the comparison of subsystems of hybrid honeypots.

TABLE III
COMPARISON OF SUBSYSTEMS OF HYBRID HONEYPOTS

| Hybrid honeypot | Frontend | Controller | Backend |
|---|---|---|---|
| Bailey's system | Honeyd | A central controller | VMware VM |
| Artail's system | Honeyd | Honeywall | Physical machine |
| GQ | Network telescopes | Click based router | VMware ESX |
| SweetBait | Honeyd+ honeycomb | - | QEMU based Argos |
| Honeybrid | Honeyd | Honeybrid gateway | VMware VM |
| SGNET | Honeyd+ ScriptGen | SGNET gateway | QEMU based Argos |
| Li's system | Spamtrap+ Phoneybot+ Phoneytoken | - | Phoneypot |
| VMI-Honeymon | Dionaea | Honeybrid gateway | Xen VM |
| IoTPOT | Frontend responder | - | QEMU based IoTBox |
| Hyhoneydv6 | Honeydv6 | - | QEMU VM |

We discover that Honeyd takes the role of the frontend in most hybrid honeypots. The wide applications of Honeyd probably attribute to its advantages of lightweight design, distributed appearance, programmable artifice and dynamic feature. Honeyd is a virtual LIH framework which can deploy multiple decoys concurrently following a certain network topology. Though it can only emulate LIHs, it still has some advantages: 1) based on the OS fingerprinting database of Nmap, it can fabricate decoys with almost all the common OS fingerprints; 2) the users can implement their own service fake response through python programming for capturing data. Honeyd may emulate a service in such a way that it can actually collect more information than a HIH would; 3) it can dynamically reconfigure the decoys by using a doorway called Honeydctl to communicate the inner workings of Honeyd.

Also, a number of controllers have been developed that mainly provide the security functions, e.g. inbound data filtering, outbound data containment, dynamic configuration, etc. Most of them are based on programable frameworks, e.g. GQ gateway is based on the Click, and Honeybrid gateway is based on the libnetfilter. These programmable frameworks allow the developers to implement their own data control functions according to the specific requirements.

It is also obvious that most hybrid honeypots use virtual machines to deploy their backends. We can see that the most popular hypervisors are Xen and QEMU. Many solutions of dynamic configuration and virtual memory introspection have been proposed based on them. With the evolvement of QEMU-KVM, we can also foresee that the KVM will be in charge of deploying HIHs for the backends. The detail analysis of virtualization enhancing honeypot development will be described in the next subsection.

## *B. Virtual honeypots*

In essential, the development of honeypots highly relies on the progress of virtualization technology. Virtual honeypots provide several valuable advantages: ease of maintenance, dynamic configuration and anti-detection.

The virtualization leads to the ease of maintenance. First, using virtualization technology, one physical machine can host multiple virtual honeypots simultaneously, which can highly improve resource efficiency. Second, the time-consuming of honeypot large-scale deployment has greatly decreased by using the virtualization techniques that only needs several minutes, while it often takes several hours to deploy honeypots by using physical machines, e.g. the physical honeypot designed by Cliff Stoll in 1986 [14].

The virtualization also facilitates the capability of dynamic configuration. The dynamic configuration is often used to reduce the responsive time to specific event. As stated, the dynamic honeypots can be used to clone the production systems and synchronize the changes of the production ones timely, and also, it can be used to investigate intrusion by modifying its own state according to the attacking research requirement. For example, the dynamic configuration can facilitate the redirection containment by redirecting the traffic back to a dynamically created honeypot, in order to control the outbound attack rather than using the brute tarpitting approach. This function also improves the capability of anti-detection, which will be described in the next paragraph.

Anti-detection is aimed to avoid the honeypot being detected. The virtualization technology provides several ways to hide both the decoy and the security program. On one hand, for hiding the security program, firstly, the virtual machine memory introspection facilitates the "out-of-the-box" monitoring approach, which improves the stealthy monitoring capability for HIHs. Secondly, as mentioned above, because the brute tarpitting approach is easy to be detected by a skilled adversary, dynamic honeypot systems often redirect the outbound traffic back into the honeynet for anti-detection. On the other hand, for limited-functional honeypots, the anti-detecting will focus on camouflaging the fact that the decoy is a honeypot. For example, due to the link latency of Honeyd based decoy can be used to detect the fact, Fu et al. [90] improved the Honeyd through reducing the link latency in order to camouflage the Honeyd based decoy. Additionally, if a decoy has been detected, the inbound traffic rate to the honeypot will be reduced [91], so in this case, the system can redeploy the decoy for performing anti-detection.

### C. Special purpose honeypots

As stated, an increasing number of special purpose honeypots [86], [47], [48], [49] have been developed. Firstly, both the independent honeypots and the cooperative honeypots are focusing on developing specific attacked resource oriented honeypots. Because these honeypots focus on fully emulating one type of information resource so that they have a higher opportunity to get fine-grained data. With the rapid growth of cyberspace, the SCADA/ICS and IoT facilities are also faced with various cyber threats every day. As stated, corresponding honeypots for those areas are also developed. Thus, the tendency of honeypot development is closely related to the industrial production as well. Secondly, research honeypots, particularly, for anomaly-detection and attack profiling, have become increasingly numerous, which highly relies on the cutting-edge technologies of other sciences, such as machine learning, big data analysis, etc. That is because honeypot is a rapidly developing interdisciplinary field that often needs expertise over various disciplines.

## V. Conclusion

As an emerging and rapidly developing interdisciplinary, honeypot has become a hot research in the field of computer and network security. From a variety of honeypot systems, we captured two common essential elements, decoy and security program. We discovered the decoupling trend of the two elements' organization. Though the capability of a honeypot has become increasingly augmented, the coupling trend has been evolving from tight to loose in order to reduce the risk that a change made within one components will create unanticipated changes within others.

Furthermore, based on the core concept, a novel D-P based taxonomy of honeypot systems is proposed, which can help to distinctly investigate the honeypot systems and techniques. Thanks to the taxonomy, we specialised in various decoy features and reviewed a number of honeypot related technologies in interdisciplinary and cutting-edge sciences. Broadly speaking, current honeypot development includes two vital parts: independent honeypots and cooperative honeypots. So, on one hand, owing to the advantages of lightweight design, low-cost development, easy management, resource efficiency, etc, the independent honeypots have a steady development and various application scenarios, e.g. numerous specific resource oriented honeypots have emerged as independent honeypot software. On the other hand, the cooperative honeypots can not only provide broader views due to the distributed and cooperative deployment in different network domains, but also create opportunities of early network anomaly detection, attack correlation, and global network status inference. Also, the cooperative ones own robustness, reliability, reusability, and understandability because of their decoupling feature.

All in all, though current honeypots have been evolving to be increasingly complex and powerful, the decoy and security program are the two fundamental elements, which originate all the development in this important area. Therefore, this work can help security researchers to gain insight into the honeypot research area and explore the design and application space for the future honeypot systems.

## Acknowledgment

The authors would like to appreciate Professor David Chadwick at University of Kent, Canterbury, UK. He conducted a thorough proofreading that improves the quality of the whole paper.

## References

[1] S. Brandes, "The newest warfighting domain: Cyberspace," *Synesis: A Journal of Science, Technology, Ethics, and Policy*, vol. 4, pp. G90–95, 2013.
[2] "Internet security threat report," Symantec, Technical Report, 2016.
[3] G. J. Rattray, "An environmental approach to understanding cyberpower," *Cyberpower and National Security*, pp. 253–274, 2009.
[4] S. Peisert, M. Bishop, and K. Marzullo, "What do firewalls protect? an empirical study of firewalls, vulnerabilities, and attacks," UC Davis CS, Technical Report CSE-2010-8, 2010.
[5] K. G. Anagnostakis, S. Sidiroglou, P. Akritidis, K. Xinidis, E. P. Markatos, and A. D. Keromytis, "Detecting targeted attacks using shadow honeypots," in *Usenix Security*, 2005.
[6] W. Fan, K. Lwakatare, and R. Rong, "Social engineering: I-e based model of human weakness for attack and defense investigations," *International Journal of Computer Network and Information Security*, vol. 9, no. 1, pp. 1–11, 2017.
[7] V. Mateos, V. A. Villagrá, F. Romero, and J. Berrocal, "Definition of response metrics for an ontology-based automated intrusion response systems," *Computers & Electrical Engineering*, vol. 38, no. 5, pp. 1102–1114, 2012, special issue on Recent Advances in Security and Privacy in Distributed Communications and Image processing.
[8] G. Portokalidis, A. Slowinska, and H. Bos, "Argos: An emulator for fingerprinting zero-day attacks for advertised honeypots with automatic signature generation," in *Proceedings of the 1st ACM SIGOPS/EuroSys European Conference on Computer Systems 2006*, ser. EuroSys '06. New York, NY, USA: ACM, 2006, pp. 15–27.
[9] H. Artail, H. Safa, M. Sraj, I. Kuwatly, and Z. Al-Masri, "A hybrid honeypot framework for improving intrusion detection systems in protecting organizational networks," *Comput. Secur.*, vol. 25, no. 4, pp. 274–288, Jun. 2006.
[10] M. Bailey, E. Cooke, D. Watson, F. Jahanian, and N. Provos, "A hybrid honeypot architecture for scalable network monitoring," *Technical Report CSE-TR-499-04, University of Michigan*, 2004.
[11] M. Nawrocki, M. Wählisch, C. Schmidt, T. C. andKeil, and J. Schönfelder, "A survey on honeypot software and data analysis," *ArXiv e-prints*, Aug. 2016.
[12] L. Spitzner, "The honeynet project: trapping the hackers," *IEEE Security Privacy*, vol. 1, no. 2, pp. 15–23, Mar 2003.
[13] T. K. Lengyel, J. Neumann, S. Maresca, B. D. Payne, and A. Kiayias, "Virtual machine introspection in a hybrid honeypot architecture," in *Presented as part of the 5th Workshop on Cyber Security Experimentation and Test*. Berkeley, CA: USENIX, 2012.
[14] C. Stoll, *The Cuckoo's Egg: Tracking a Spy Through the Maze of Computer Espionage*. Gallery Books, 2000.
[15] B. Cheswick, "An evening with berferd in which a cracker is lured, endured, and studied," in *In Proc. Winter USENIX Conference*, 1992, pp. 163–174.
[16] L. Spitzner, "Honeypots: catching the insider threat," in *Computer Security Applications Conference, 2003. Proceedings. 19th Annual*, Dec 2003, pp. 170–179.
[17] N. Provos, "A virtual honeypot framework," in *Proceedings of the 13th Conference on USENIX Security Symposium (SSYM'04)*, Berkeley, CA, USA, 2004, pp. 1–14.
[18] "Dionaea - catched bugs," Nov. 2011. [Online]. Available: http://dionaea.carnivore.it/
[19] M. Oosterhof, "Cowrie - active kippo fork," July 2015. [Online]. Available: http://www.micheloosterhof.com/cowrie/
[20] CuckooFoundation, "Cuckoo," Oct. 2014. [Online]. Available: http://www.cuckoosandbox.org/
[21] D. Moore, C. Shannon, G. M. Voelker, and S. Savage, "Network Telescopes: Technical Report," University of California, San Diego, Tech. Rep., July 2004.

[22] T. CYMRU, “The darknet project,” Jul. 2015. [Online]. Available: http://www.team-cymru.org/darknet.html

[23] D. Song, R. Malan, and R. Stone, “A snapshot of global internet worm activity,” in *In Proc. first conference on computer security incident handling and response*, Jun. 2002.

[24] M. Bailey, E. Cooke, F. Jahanian, J. Nazario, and D. Watson, “The internet motion sensor: A distributed blackhole monitoring system,” in *In Proceedings of Network and Distributed System Security Symposium (NDSS 05*, 2005, pp. 167–179.

[25] V. Yegneswaran, P. Barford, and D. Plonka, “On the design and use of internet sinks for network abuse monitoring,” in *Recent Advances in Intrusion Detection*, ser. Lecture Notes in Computer Science, E. Jonsson, A. Valdes, and M. Almgren, Eds. Springer Berlin Heidelberg, 2004, vol. 3224, pp. 146–165.

[26] M. Vrable, J. Ma, J. Chen, D. Moore, E. Vandekieft, A. Snoeren, G. Voelker, and S. Savage, “Scalability, Fidelity and Containment in the Potemkin Virtual Honeyfarm,” *ACM Symposium on Operating System Principles (SOSP)*, vol. 39, no. 5, pp. 148–162, Oct 2005.

[27] “Know your enemy: Honeynets,” May 2006. [Online]. Available: http://old.honeynet.org/papers/honeynet/

[28] L. Spitzner, “Specter: A commercial honeypot solution for windows,” 2003. [Online]. Available: http://www.symantec.com/connect/articles/specter-commercial-honeypot-solution-windows/

[29] L. Rist, “Glastopf project,” 2009. [Online]. Available: http://glastopf.org/

[30] S. Poeplau and J. Gassen, “A honeypot for arbitrary malware on usb storage devices,” in *2012 7th International Conference on Risks and Security of Internet and Systems (CRiSIS)*, Oct 2012, pp. 1–8.

[31] N. Provos and T. Holz, *Virtual honeypots: from botnet tracking to intrusion detection*, 1st ed. Addison Wesley, Jul. 2007.

[32] F. Galán, D. Fernández, W. Fuertes, M. Gómez, and J. E. López de Vergara, “Scenario-based virtual network infrastructure management in research and educational testbeds with vnuml,” *annals of telecommunications - annales des télécommunications*, vol. 64, no. 5, pp. 305–323, 2009.

[33] L. K. Yan, “Virtual honeynets revisited,” in *Information Assurance Workshop, 2005. IAW ’05. Proceedings from the Sixth Annual IEEE SMC*, June 2005, pp. 232–239.

[34] F. Abbasi and R. Harris, “Experiences with a generation iii virtual honeynet,” in *Telecommunication Networks and Applications Conference (ATNAC), 2009 Australasian*, Nov 2009, pp. 1–6.

[35] A. Capalik, “Next-generation honeynet technology with real-time forensics for u.s. defense,” in *Military Communications Conference, 2007. MILCOM 2007. IEEE*, Oct 2007, pp. 1–7.

[36] N. Memari, K. Samsudin, and S. Hashim, “Towards virtual honeynet based on lxc virtualization,” in *Region 10 Symposium, 2014 IEEE*, April 2014, pp. 496–501.

[37] P. Kasza, “Creating honeypots using docker,” 2015. [Online]. Available: https://www.itinsight.hu/blog/posts/2015-05-04-creating-honeypots-using-docker.html

[38] F. Galán and D. Fernández, “Use of vnuml in virtual honeynets deployment,” *IX Reunión Española sobre Criptología y Seguridad de la Información (RECSI), Barcelona (Spain)*, 2006.

[39] F. Stumpf, A. Görlach, F. Homann, and L. Brückner, “Nose-building virtual honeynets made easy,” in *Proc. of the 12th Intl Linux System Technology Conference (Linux-Kongress 05), GUUG*, 2005, pp. 1664–1669.

[40] D. Fernandez, A. Cordero, J. Somavilla, J. Rodriguez, A. Corchero, L. Tarrafeta, and F. Galan, “Distributed virtual scenarios over multi-host linux environments,” in *Systems and Virtualization Management (SVM), 2011 5th International DMTF Academic Alliance Workshop on*, Oct 2011, pp. 1–8.

[41] W. Fan, D. Fernndez, and Z. Du, “Versatile virtual honeynet management framework,” *IET Information Security*, vol. 11, no. 1, pp. 38–45, March 2016.

[42] W. Chin, E. Markatos, S. Antonatos, and S. Ioannidis, “Honeylab: Large-scale honeypot deployment and resource sharing,” in *Network and System Security, 2009. NSS ’09. Third International Conference on*, Oct 2009, pp. 381–388.

[43] B. Sobesto, M. Cukier, M. Hiltunen, D. Kormann, G. Vesonder, and R. Berthier, “Darknoc: Dashboard for honeypot management,” in *Proceedings of the 25th International Conference on Large Installation System Administration*, ser. LISA’11. Berkeley, CA, USA: USENIX Association, 2011, pp. 16–16.

[44] W. Han, Z. Zhao, A. Doupé, and G.-J. Ahn, “Honeymix: Toward sdn-based intelligent honeynet,” in *Proceedings of the 2016 ACM International Workshop on Security in Software Defined Networks & Network Function Virtualization*, ser. SDN-NFV Security ’16. New York, NY, USA: ACM, 2016, pp. 1–6.

[45] R. do Carmo, M. Nassar, and O. Festor, “Artemisa: An open-source honeypot back-end to support security in voip domains,” in *12th IFIP/IEEE International Symposium on Integrated Network Management (IM 2011) and Workshops*, May 2011, pp. 361–368.

[46] A. Podhradsky, C. Casey, and P. Ceretti, “The bluetooth honeypot project: Measuring and managing bluetooth risks in the workplace,” *Int. J. Interdiscip. Telecommun. Netw.*, vol. 4, no. 3, pp. 1–22, Jul. 2012.

[47] L. Rist, J. Vestergaard, D. Haslinger, A. Pasquale, and J. Smith, “Conpot ics/scada honeypot,” November 2013. [Online]. Available: http://conpot.org/

[48] Y. M. P. Pa, S. Suzuki, K. Yoshioka, T. Matsumoto, T. Kasama, and C. Rossow, “Iotpot: Analysing the rise of iot compromises,” in *9th USENIX Workshop on Offensive Technologies (WOOT 15)*. Washington, D.C.: USENIX Association, Aug. 2015.

[49] S. Schindler, B. Schnor, and T. Scheffler, “Hyhoneydv6: A hybrid honeypot architecture for ipv6 networks,” *International Journal of Intelligent Computing Research*, vol. 6, 2015.

[50] W. Cui, V. Paxson, and N. C. Weaver, “Gq: Realizing a system to catch worms in a quarter million places,” University of California, Berkeley, CA, Technical Report TR-06-004, 2006.

[51] “Know your enemy: Sebek, a kernel based data capture tool,” nov 2003. [Online]. Available: http://old.honeynet.org/papers/sebek.pdf

[52] “Know your tools: Qebek - conceal the monitoring,” nov 2010. [Online]. Available: http://www.honeynet.org/papers/KYT_qebek

[53] C. Willems, T. Holz, and F. Freiling, “Toward automated dynamic malware analysis using cwsandbox,” *IEEE Security Privacy*, vol. 5, no. 2, pp. 32–39, March 2007.

[54] X. Jiang and X. Wang, “out-of-the-box monitoring of vm-based high-interaction honeypots,” in *Recent Advances in Intrusion Detection*, ser. Lecture Notes in Computer Science, C. Kruegel, R. Lippmann, and A. Clark, Eds. Springer Berlin Heidelberg, 2007, vol. 4637, pp. 198–218.

[55] LibVMIProject, “Libvmi,” 2015. [Online]. Available: http://libvmi.com/

[56] T. Garfinkel and M. Rosenblum, “A virtual machine introspection based architecture for intrusion detection,” in *In Proc. Network and Distributed Systems Security Symposium*, 2003, pp. 191–206.

[57] B. D. Payne, M. Carbone, M. Sharif, and W. Lee, “Lares: An architecture for secure active monitoring using virtualization,” in *2008 IEEE Symposium on Security and Privacy (sp 2008)*, May 2008, pp. 233–247.

[58] X. Jiang, X. Wang, and D. Xu, “Stealthy malware detection through vmm-based ”out-of-the-box” semantic view reconstruction,” in *Proceedings of the 14th ACM Conference on Computer and Communications Security*, ser. CCS ’07. New York, NY, USA: ACM, 2007, pp. 128–138.

[59] J. Pfoh, C. Schneider, and C. Eckert, “Nitro: Hardware-based system call tracing for virtual machines,” in *Advances in Information and Computer Security*, ser. Lecture Notes in Computer Science, T. Iwata and M. Nishigaki, Eds. Springer Berlin Heidelberg, 2011, vol. 7038, pp. 96–112.

[60] D. Srinivasan and X. Jiang, “Time-traveling forensic analysis of vm-based high-interaction honeypots,” in *Proceedings of the 7th International Conference on Security and Privacy in Communication Networks*, ser. SecureComm2011, London, UK, September 2011.

[61] B. Dolan-Gavitt, T. Leek, M. Zhivich, J. Giffin, and W. Lee, “Virtuoso: Narrowing the semantic gap in virtual machine introspection,” in *2011 IEEE Symposium on Security and Privacy*, May 2011, pp. 297–312.

[62] T. K. Lengyel, S. Maresca, B. D. Payne, G. D. Webster, S. Vogl, and A. Kiayias, “Scalability, fidelity and stealth in the drakvuf dynamic malware analysis system,” in *Proceedings of the 30th Annual Computer Security Applications Conference*, ser. ACSAC ’14. New York, NY, USA: ACM, 2014, pp. 386–395.

[63] R. Pang, V. Yegneswaran, P. Barford, V. Paxson, and L. Peterson, “Characteristics of internet background radiation,” in *Proceedings of the 4th ACM SIGCOMM conference on Internet measurement*. ACM, 2004, pp. 27–40.

[64] M. Bailey, E. Cooke, F. Jahanian, N. Provos, K. Rosaen, and D. Watson, “Data reduction for the scalable automated analysis of distributed darknet traffic,” in *Proceedings of the 5th ACM SIGCOMM conference on Internet Measurement*. USENIX Association, 2005, pp. 21–21.

[65] G. Portokalidis and H. Bos, “Sweetbait: Zero-hour worm detection and containment using low-and high-interaction honeypots,” *Computer Networks*, vol. 51, no. 5, pp. 1256–1274, 2007.

[66] W. Cui, V. Paxson, N. C. Weaver, and Y. H. Katz, "Protocol-independent adaptive replay of application dialog," in *In The 13th Annual Network and Distributed System Security Symposium (NDSS)*, 2006.
[67] X. Jiang and D. Xu, "Collapsar: A vm-based architecture for network attack detention center." in *USENIX Security Symposium*, 2004, pp. 15–28.
[68] "Know your enemy: Honeywall cdrom," May 2005. [Online]. Available: http://old.honeynet.org/papers/cdrom/
[69] É. Alata, I. Alberdi, V. Nicomette, P. Owezarski, and M. Kaâniche, "Internet attacks monitoring with dynamic connection redirection mechanisms," *Journal in Computer Virology*, vol. 4, no. 2, pp. 127–136, 2008.
[70] J. Newsome and D. Song, "Dynamic taint analysis for automatic detection, analysis, and signature generation of exploits on commodity software," in *Proceedings of the 12th Annual Network and Distributed System Security Symposium (NDSS '05)*, 2005.
[71] C. Kreibich and J. Crowcroft, "Honeycomb: Creating intrusion detection signatures using honeypots," *SIGCOMM Comput. Commun. Rev.*, vol. 34, no. 1, pp. 51–56, Jan. 2004.
[72] A. Javaid, Q. Niyaz, W. Sun, and M. Alam, "A deep learning approach for network intrusion detection system," in *Proceedings of the 9th EAI International Conference on Bio-inspired Information and Communications Technologies (Formerly BIONETICS)*, ser. BICT'15. ICST, Brussels, Belgium, Belgium: ICST (Institute for Computer Sciences, Social-Informatics and Telecommunications Engineering), 2016, pp. 21–26.
[73] R. Sekar, A. Gupta, J. Frullo, T. Shanbhag, A. Tiwari, H. Yang, and S. Zhou, "Specification-based anomaly detection: A new approach for detecting network intrusions," in *Proceedings of the 9th ACM Conference on Computer and Communications Security*, ser. CCS '02, Washington, DC, USA, 2002, pp. 265–274.
[74] E. Kohler, R. Morris, B. Chen, J. Jannotti, and M. F. Kaashoek, "The click modular router," *ACM Trans. Comput. Syst.*, vol. 18, no. 3, pp. 263–297, Aug. 2000.
[75] C. Yoon, T. Park, S. Lee, H. Kang, S. Shin, and Z. Zhang, "Enabling security functions with sdn: A feasibility study," *Computer Networks*, vol. 85, no. C, pp. 19 – 35, 2015.
[76] R. Berthier and M. Cukier, "Honeybrid: A hybrid honeypot architecture," 2008.
[77] H. Welte and P. N. Ayuso, "The libnetfilter_queue project," 2014. [Online]. Available: http://www.netfilter.org/projects/libnetfilter_queue/
[78] Y.-D. Lin, T.-B. Shih, Y.-S. Wu, and Y.-C. Lai, "Secure and transparent network traffic replay, redirect, and relay in a dynamic malware analysis environment," *Security and Communication Networks*, vol. 7, no. 3, pp. 626–640, 2014.
[79] T. Lengyel, J. Neumann, S. Maresca, and A. Kiayias, "Towards hybrid honeynets via virtual machine introspection and cloning," in *Network and System Security*, ser. Lecture Notes in Computer Science, J. Lopez, X. Huang, and R. Sandhu, Eds. Springer Berlin Heidelberg, 2013, vol. 7873, pp. 164–177.
[80] W. Fan and D. Fernández, "A novel sdn based stealthy tcp connection handover mechanism for hybrid honeypot systems," in *Proceedings of IEEE 3rd Conference on Network Softwarization (NetSoft2017)*, Bologna, Italy, July 2017.
[81] C. Hecker and B. Hay, "Automated honeynet deployment for dynamic network environment," in *System Sciences (HICSS), 2013 46th Hawaii International Conference on*, Jan 2013, pp. 4880–4889.
[82] M. Zalewski, "pof v3," 2012-2014. [Online]. Available: http://lcamtuf.coredump.cx/p0f3/
[83] G. Lyon, "Namp," 2015. [Online]. Available: http://nmap.org/
[84] R. McGrew and R. B. Vaughn JR, "Experiences with honeypot systems: Development, deployment, and analysis," in *Proceedings of the 39th Annual Hawaii International Conference on System Sciences (HICSS'06)*, vol. 9, Jan 2006, p. 220a.
[85] M. Egele, T. Scholte, E. Kirda, and C. Kruegel, "A survey on automated dynamic malware-analysis techniques and tools," *ACM Comput. Surv.*, vol. 44, no. 2, pp. 6:1–6:42, Mar. 2008.
[86] K. Rieck, P. Trinius, C. Willems, and T. Holz, "Automatic analysis of malware behavior using machine learning," *J. Comput. Secur.*, vol. 19, no. 4, pp. 639–668, Dec. 2011.
[87] "Know your enemy: Genii honeynets," May 2005. [Online]. Available: http://old.honeynet.org/papers/gen2/
[88] C. Leita and M. Dacier, "Sgnet: A worldwide deployable framework to support the analysis of malware threat models," in *Dependable Computing Conference, 2008. EDCC 2008. Seventh European*, May 2008, pp. 99–109.
[89] S. Li and R. Schmitz, "A novel anti-phishing framework based on honeypots," in *2009 eCrime Researchers Summit*, Sept 2009, pp. 1–13.
[90] X. Fu, B. Graham, D. Cheng, R. Bettati, and W. Zhao, "Camouflaging virtual honeypots," in *In Texas A&M University*, 2005.
[91] H. Wang and Q. Chen, "Dynamic deploying distributed low-interaction honeynet," *Journal of Computers*, vol. 7, no. 3, 2012.

**Wenjun Fan** is a Ph.d. candidate at the Department of Telematics Engineering at Technical University of Madrid (UPM) since 2011. His research interests are cyber security, SDN and NFV, cloud computing and adversarial machine learning.

**Zhihui Du** received the BE degree in 1992 in Computer Department from Tianjian University. He received the MS and PhD degrees in computer science, respectively, in 1995 and 1998, from Peking University. From 1998 to 2000, he worked at Tsinghua University as a postdoctoral researcher. Since 2001, he is working at Tsinghua University as an associate professor in the Department of Computer Science and Technology. His research areas include high performance computing and grid computing.

**David Fernández** is an associate professor of computer networks at Technical University of Madrid (UPM). He received a telecommunications engineering M.Sc. degree in 1988 and a Ph.D. in telematics engineering in 1993, both from Technical University of Madrid. His research interests are computer-supported cooperative work, advanced Internet protocols, and network virtualization.

**Víctor A. Villagrá** is an associate professor in telematics engineering at the UPM since 1992. His research interests include Network Management, Advanced Services Design and Network Security.